\documentclass[epjST]{svjour}
\usepackage{graphicx}
\usepackage{amsmath}
\usepackage{amssymb}
\usepackage{dcolumn}
\usepackage[english]{babel}
\usepackage[numbers,square,comma,sort&compress]{natbib}
\begin{document}
\bibliographystyle{apsrev}
\graphicspath {{./figures/}}
\makeatletter
\def\input@path{{./figures/}}
\makeatother
\newcolumntype{d}[1]{D{.}{.}{#1}}

\title{Exploring first-order phase transitions with population annealing}

\author{
  Lev Yu. Barash\inst{1,2}
  \and Martin Weigel\inst{3}\fnmsep\thanks{\email{martin.weigel@coventry.ac.uk}}
  \and Lev N. Shchur\inst{1,2,4}
  \and Wolfhard Janke\inst{5}
}
\institute{
  Science Center in Chernogolovka, 142432 Chernogolovka, Russia
  \and Landau Institute for Theoretical Physics, 142432 Chernogolovka, Russia
  \and Applied Mathematics Research Centre, Coventry University, Coventry CV1 5FB, United Kingdom
  \and National Research University Higher School of Economics, 101000 Moscow, Russia
  \and Institut f\"ur Theoretische Physik, Universit\"at Leipzig, Postfach 100\,920, D-04009 Leipzig, Germany
}

\abstract{
  Population annealing is a hybrid of sequential and Markov chain Monte Carlo methods
  geared towards the efficient parallel simulation of systems with complex
  free-energy landscapes. Systems with first-order phase transitions are among the
  problems in computational physics that are difficult to tackle with standard
  methods such as local-update simulations in the canonical ensemble, for example
  with the Metropolis algorithm. It is hence interesting to see whether such
  transitions can be more easily studied using population annealing. We report here
  our preliminary observations from population annealing runs for the two-dimensional
  Potts model with $q > 4$, where it undergoes a first-order transition.
}

\maketitle

\section{Introduction}

Monte Carlo simulations are an indispensable tool for studies of a wide range of
problems in statistical physics, including magnetic systems and other models on
lattices as well as continuum models for polymers or colloids
\cite{binder:book2}. While after 50 years of research the toolbox of simulational
methods is quite well equipped with a rather diverse set of techniques, the vast
majority belong to the kingdom of Markov chain approaches. Fundamentally different
schemes such as sequential Monte Carlo \cite{doucet:13} have received significantly
less attention in this field (see, however, Ref.~\cite{grassberger:02a}). Population
annealing (PA) \cite{iba:01,hukushima:03} is a technique combining elements of Markov
chain and sequential Monte Carlo that has received relatively little attention to
date \cite{machta:10a,machta:11,wang:15a}.

Since about 2005 the race towards higher and higher clock frequencies of CPUs and the
resulting constant increase of the performance available from serial codes have come
to end. High-performance computing has hence arrived in the era of massive
parallelism, where additional computational power is essentially only available from
a further multiplication of parallel computational cores \cite{asanovic:06}. This
also led to a widespread application of hardware accelerators such as graphics
processing units (GPUs) or Intel's Xeon Phi devices, which currently feature several
hundreds up to several thousands of cores per device. To be able to tap into this
massively parallel computational power one needs parallel algorithms that scale well
with the number of cores \cite{owens:08}. This is not one of the main strengths of
Markov chain Monte Carlo (MCMC), which is inherently sequential, and parallelism can
only be employed by sub-dividing the work in the updating step (domain decomposition)
or by running multiple chains in parallel. The parallelism in the former approach is
limited by the size of systems studied, while the latter has asymptotically vanishing
efficiency as a larger and larger fraction of time needs to be spent on
equilibration. This is where PA comes to the rescue. In PA one starts from a
population of uncorrelated, random configurations at infinite temperature that are
propagated down to low temperatures according to a well-defined stochastic protocol.
For a population of size $R$ statistical errors decrease like $1/\sqrt{R}$ and bias
as $1/R$ \cite{machta:11,wang:15a}. The size of populations is mostly limited only by
the available memory, but since memory is typically expected to scale with the number
of cores this is not a real problem. The approach hence has theoretically excellent
scaling properties, which are also borne out very well in practical implementations,
for example on GPU \cite{barash:16}.

An important application field for computer simulation studies in statistical physics
are phase transitions and critical phenomena. While a lot of effort on the
theoretical and computational side has been invested in the understanding of systems
with continuous transitions, the vast majority of phase transitions in nature is of
first order. They are characterized by the coexistence of two (or more) phases at the
transition point as well as the phenomenon of metastability, i.e., the system remains
in its present phase when crossing the transition point \cite{binder:87}. These
effects are accompanied by discontinuities in observable quantities such as the
internal energy or magnetization across the transition, as well as dynamic effects
such as hysteresis. In contrast to second-order transitions the correlation length
remains finite. These features result in particular challenges for simulations of
systems undergoing first-order transitions, including an exponential slowing down of
the dynamics connecting the two phases due to a region of strongly suppressed states
\cite{janke:03}. Well known rather efficient simulation methods for this situation
are the multicanonical approach \cite{berg:92b} and derived techniques such as
Wang-Landau sampling \cite{wang:01a}. While population annealing has been used for
simulations of spin-glass systems \cite{wang:14,wang:15a,wang:15b} and the Ising
model \cite{weigel:17}, as well as for finding ground states of frustrated systems
\cite{wang:15,borovsky:16}, its behavior for systems undergoing first-order phase
transitions has not been studied to date. We report here some preliminary results
demonstrating the behavior of the PA algorithm for simulations in the first-order
regime of the $q$-states Potts model in two dimensions \cite{wu:82a}.

The rest of the paper is organized as follows. In Sec.~\ref{sec:algorithm} we
summarize the PA algorithm, while in Sec.~\ref{sec-potts} we introduce the Potts
model and the relevant observables considered here. In Sec.~\ref{sec:widths} we
report some properties of the distribution of energies and magnetizations in the
population in the vicinity of a first-order transition. In Sec.~\ref{sec:hysteresis}
we show that PA is affected by hysteresis effects for discontinuous
transitions. Section \ref{sec:integration} is devoted to the illustration of a method
of using the free-energy estimator provided by PA to determine the location of the
transition point. Finally, Sec.~\ref{sec:conclusions} contains our conclusions.

\section{The population annealing algorithm}
\label{sec:algorithm}

Population annealing is a weighted sequential algorithm that performs a temperature
sweep of a population of configurations (replicas) of the system under consideration
\cite{iba:01,hukushima:03,machta:10a}. At each temperature step the population is
resampled from the current distribution at inverse temperature $\beta=1/k_B T$
according to the probability distribution of energies at the target temperature
$\beta+\Delta\beta$ that is estimated by reweighting. If the initial population is in
equilibrium, which can be easily achieved by starting with random configurations
produced by simple sampling at infinite temperature $\beta_0 = 0$, this procedure
keeps the ensemble at equilibrium at all subsequent temperatures. In practice,
however, the resampling leads to an exponential decline of diversity in the
population, and in order to ensure fair sampling one needs to augment the procedure
with further updates on the individual replicas that will typically be chosen
according to a Markov chain scheme. In detail, the algorithm comprises the following
steps:
\begin{enumerate}
\item Set up an equilibrium ensemble of $R = R_0$ independent copies (replicas) of
  the system at inverse temperature $\beta_0$. Often $\beta_0 = 0$, where this
  can be easily achieved.
\item To create an approximately equilibrated sample at $\beta_i > \beta_{i-1}$,
  resample configurations with their relative Boltzmann weight
  $\tau_i(E_j) = \exp[-(\beta_i-\beta_{i-1})E_j]/Q_i$, where
  $Q_i = \sum_j \exp[-(\beta_i-\beta_{i-1})E_j]/R_{i-1}$.
\item Update each replica by $\theta$ rounds of an MCMC algorithm at inverse
  temperature $\beta_i$.
\item Calculate estimates for observable quantities ${\cal O}$ as population averages
  $\sum_j {\cal O}_j/R_i$.
\item Goto step 2 unless the target temperature $\beta_{K-1}$ has been reached.
\end{enumerate}
While there is no theoretical restriction on the (inverse) temperature protocol
$\beta_0,\ldots$, $\beta_{K-1}$ to be used, we focus here on the simplest choice of
constant steps, $\beta_i = \beta_{i-1} + \Delta\beta$. The resampling proportional to
$\tau_i(E_j)$ needs to take a normalization into account to ensure that the
population size stays close to $R$. One possible implementation which is used here is
to determine the number $r_j^i$ of copies of replica $j$ at temperature $\beta_i$ by
drawing a random number from a Poisson distribution,
\begin{equation}
  \label{eq:poisson}
  r_j^i \sim \mathrm{Pois}\left[(R/R_{i-1})\tau_i(E_j)\right].
\end{equation}
The new population size is then $R_i = \sum_j r^i_j$. The equilibration sweeps in
step 3 can be chosen freely from any importance sampling algorithm. Here we use
simple Metropolis single-spin flip updates.

A specialty of the PA approach is that it provides a natural estimate of the free
energy through the expression \cite{machta:10a}
\begin{equation}
  -\beta_i F({\beta_i}) = \ln Z_{\beta_0} + \sum_{m=1}^{i} \ln Q_m,
  \label{eq:free-energy}
\end{equation}
that involves the reweighting factors $Q_m$. Here, $Z_{\beta_0}$ denotes the partition
function at $\beta_0$ which needs to be known from other sources to get absolute free
energies instead of just free-energy differences. This can be provided by explicit
calculation for instance for $\beta_0 = 0$ or $\beta_0\to\infty$ or, more generally,
through the application of high- and low-temperature expansions.

\section{Potts model and observables}
\label{sec-potts}

The Potts model is a natural generalization of the Ising model to spins with $q$
different states. The Hamiltonian in zero field is \cite{wu:82a}
\begin{equation}
{\cal H} = -J\sum_{\langle ij\rangle}\delta_{s_i,s_j},
\label{Z-PM}
\end{equation}
where the spins $s_i=1,2,...,q$ and $J > 0$ is a ferromagnetic coupling constant. We
study the model on the square lattice with nearest-neighbor interactions as indicated
by the notation $\langle ij\rangle$ and set $J=1$ to fix units. Periodic boundary
conditions are applied. For this specific setup, the transition temperature is
exactly given by the relation $\beta_t= \ln{(1{+}\sqrt{q)}}$ that follows from the
self-duality of the model \cite{baxter:73a,beffara:12}. The model shows a first-order
phase transition for $q > q_c$ and one finds that $q_c=4$ for the present setup
\cite{wu:82a}. For $q \le q_c$ the transition is continuous, with additional
logarithmic corrections directly at $q_c$.

We use population annealing with a Metropolis update on single spins in step 3 of the
algorithm described above to study the square-lattice Potts model for $q=6$, $8$,
$10$, and $20$ (as well as, for comparison, $q=3$ in the second-order regime). The
strength of the transition increases with $q$. The case $q=6$ is still relatively
weakly first-order with a correlation length $\xi\approx 160$ at the transition
point, while $q=20$ has a correlation length of $\xi\approx 3$ \cite{borgs:92}. In
contrast to regular MCMC, measurements in the PA approach are taken as ensemble
averages over the population, and we thus record
\begin{equation}
  \begin{split}
  \overline{E}(\beta_i) &= \frac{1}{R_i} \sum_{j=1}^{R_i} E_j,\\
  \overline{M}(\beta_i) &= \frac{1}{R_i} \sum_{j=1}^{R_i} M_j.
  \end{split}
  \label{eq:observables}
\end{equation}
Here, $E = {\cal H}(\{s_k\})$ is the configurational energy, and the
magnetization is defined on a finite lattice with $N=L^2$ spins via the number
$\widetilde{M}$ of spins in the most common spin orientation,
\begin{equation}
  \begin{split}
    M &= \frac{q\widetilde{M}-N}{q-1},\\
    \widetilde{M} &= \max_{1 \le \alpha \le q} \sum_{k=1}^N \delta_{s_k,\alpha}.
  \end{split}
\end{equation}

\section{Behavior of the population}
\label{sec:widths}

\begin{figure}[t]
  \begin{center}
    \includegraphics[scale=0.8]{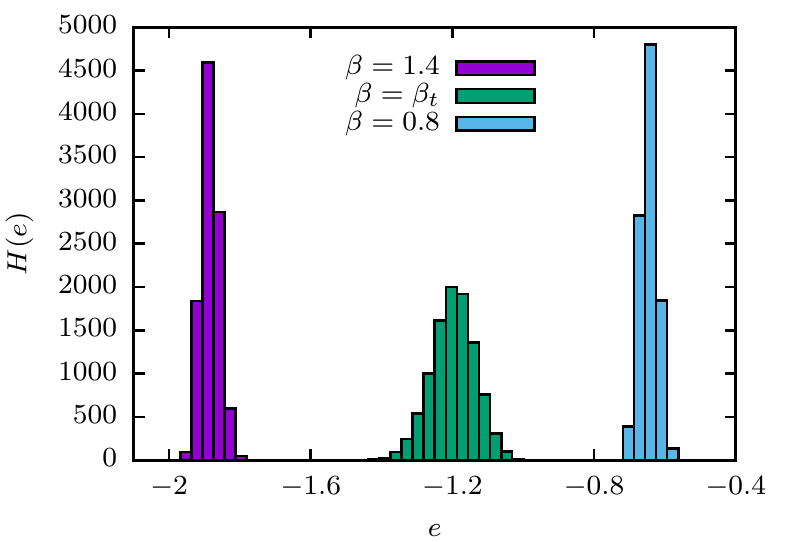}
    \includegraphics[scale=0.8]{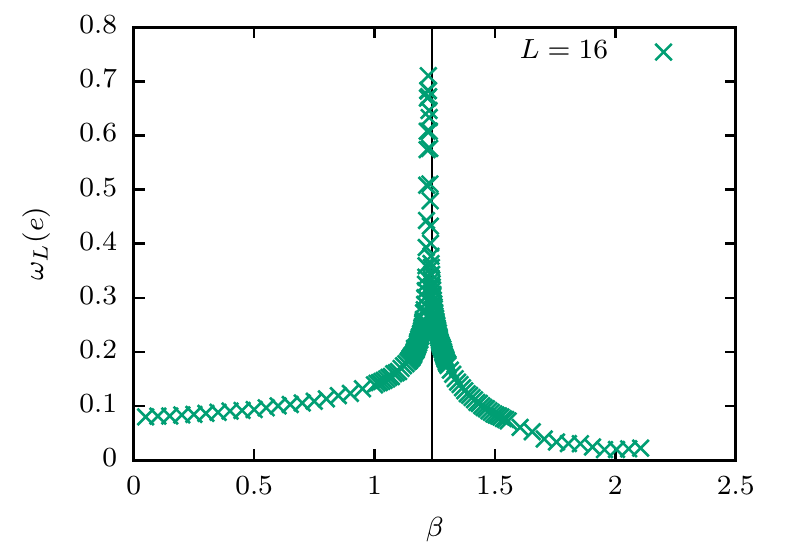}
  \end{center}
  \caption{\label{fig:histogram}%
    Left: Histograms of internal energies $e=E/N$ per spin of population members for
    a PA run for the $6$-state Potts model on a $32\times 32$ square lattice with
    periodic boundaries and parameters $R=10\,000$, $\theta=10$, $\Delta\beta = 0.01$
    at inverse temperatures $\beta = 0.8$ in the high-temperature phase,
    $\beta=\beta_t \approx 1.24$ at the transition point, and $\beta = 1.4$ in the
    ordered phase.  Right: Full width at half maximum, $\omega_L(e)$, of the energy
    distribution over the population with $R=1000$ in a PA run for the $6$-state
    model and $L=16$, $\theta=100$. We used $\Delta\beta = 1/N_\beta$ with $N_\beta =
    1000$ for $1.135 \le \beta \le 1.305$ and $\Delta\beta = 0.01$ otherwise, unless
    $\beta<0.995$ or $\beta>1.545$ in which case we used $\Delta\beta = 0.05$.
  }
\end{figure}

In a perfectly equilibrated PA simulation, the set of replicas at each temperature is
a sample from the equilibrium energy distribution. For a system in the vicinity of a
first-order transition one hence expects a rather wide distribution and, right at the
transition point, a double peak indicating the phase coexistence there
\cite{janke:03}. In the left panel of Fig.~\ref{fig:histogram} we show three
representative histograms for a PA run with $R=10\,000$ and $\Delta\beta = 0.01$ for
the $q=6$ model with $L=32$. While one clearly sees a broadening of the energy
distribution at the transition point, there is no double peak --- and for these
parameters we also do not find a double peak for any other temperature in the
vicinity of the transition point. As we will see in more detail below in
Sec.~\ref{sec:hysteresis} this is a consequence of the metastability of the
simulations.

We quantify the behavior of the histograms by systematically studying the widths
$\omega_L(e)$ and $\omega_L(m)$ of the distributions of internal energies $e=E/N$ and
magnetizations $m=M/N$ per spin, respectively, in the population. Here, the width is
defined as the full width at half maximum of the corresponding histogram, i.e., if
the maximum of the histogram is denoted as $H_{\mathrm{max}}$, it is the
distance between the two intersections of the histogram with the horizontal line at
$H_{\mathrm{max}}/2$. As is seen for an example run for $q=6$ and $L=16$ with
$R=1000$ replicas and $\theta=100$ in the right panel of Fig.~\ref{fig:histogram},
the width of the energy histogram peaks close to the transition coupling. The same
behavior is found for the magnetization histogram. We note that the widths are
related to the specific heat and magnetic susceptibility, respectively, but these are
more precisely a function of the variances of the distributions, so the relation is
merely qualitative. Due to the metastability discussed above and the fact that we use
a cooling (and not a heating) schedule, the quantities $\omega_L(e)$ and
$\omega_L(m)$ correspond to the widths of the disordered peaks only \cite{janke:03}.

In Tables \ref{tabI} to \ref{tabIII} we collect our results for the widths
$\omega_{L,\mathrm{max}}(e)$ and $\omega_{L,\mathrm{max}}(m)$ at the temperatures
$T_{\mathrm{max}}(e)$ and $T_{\mathrm{max}}(m)$, respectively, where they are
maximal. All data are averaged results from 200 independent runs. Table \ref{tabI}
shows the dependence on the number $q$ of Potts states --- and hence the strength of
the phase transition --- as well as on the system size $L$. The size dependence of
the positions of the maxima seems to be small, and it is possibly consistent with the
shift of finite-size maxima proportional to $1/N$ expected for first-order
transitions \cite{janke:03}, but we did not perform a quantitative analysis. In Table
\ref{tabII} we summarize the observed dependence of histogram widths on the number
$\theta$ of equilibration sweeps taken at each temperature. It is seen that the
widths increase with $\theta$, indicating a gradual reduction of hysteresis with
increasing $\theta$, thus ultimately revealing the double-peak nature of the energy
and magnetization histograms at the transition point. Finally, in Table \ref{tabIII}
we show the dependence of the maxima of the histogram widths on the temperature
protocol for the case of using a spacing $\Delta\beta = 1/N_\beta$ in inverse
temperature in the vicinity of the transition for different values of $N_\beta$. It
is seen that decreasing the size of temperature steps has an effect that is similar
to that of increasing $\theta$ \cite{weigel:17}.

\begin{table}[tb]
  \caption{%
    Maximal histogram widths and temperatures of maxima for the energy and
    magnetization for PA runs with $R=1000$ and $\theta=100$.  We used $N_\beta = 1000$
    in all cases. To speed up the calculations, inverse temperature steps were chosen as
    follows. $q=6$: $\Delta\beta = 1/N_\beta$ for $1.135 \le \beta \le 1.305$ and
    $\Delta\beta = 0.01$ otherwise, unless $\beta<0.995$ or $\beta>1.545$ where we used
    $\Delta\beta = 0.05$. $q=8$: $\Delta\beta = 1/N_\beta$ for $1.25 \le \beta \le 1.42$
    and $\Delta\beta = 0.01$ otherwise, unless $\beta<1.11$ or $\beta>1.66$ where we used
    $\Delta\beta = 0.05$. $q=10$: $\Delta\beta = 1/N_\beta$ for $1.305 \le \beta \le
    1.475$ and $\Delta\beta = 0.01$ otherwise, unless $\beta<1.165$ or $\beta>1.715$
    where we used $\Delta\beta = 0.05$. $q=20$: $\Delta\beta = 1/N_\beta$ for $1.735 \le
    \beta \le 1.905$ and $\Delta\beta = 0.01$ otherwise, unless $\beta<1.595$ or
    $\beta>2.145$ where we used $\Delta\beta = 0.05$.
  }
  \center
  \begin{tabular}{|c|r|d{1.3}|d{1.3}|d{1.3}|d{1.3}|}
\hline
    &  \multicolumn{1}{c|}{$L$}	& \multicolumn{1}{c|}{$\omega_{L,\mathrm{max}}(e)$} &
                  \multicolumn{1}{c|}{$T_{\mathrm{max}}(e)$} &
                  \multicolumn{1}{c|}{$\omega_{L,\mathrm{max}}(m)$} &
                  \multicolumn{1}{c|}{$T_{\mathrm{max}}(m)$} \\
\hline
     &  16	&	0.71	&	0.82	&	0.51	&	0.82 \\
     &  32	&	0.39	&	0.81	&	0.41	&	0.81 \\
$q=6$&  64	&	0.23	&	0.80	&	0.40	&	0.80 \\
     &  128	&	0.07	&	0.80	&	0.16	&	0.80 \\
     &  256	&	0.03	&	0.80	&	0.055	&	0.79 \\
\hline
     &  16	&	0.64	&	0.75	&	0.46	&	0.75 \\
     &  32	&	0.34	&	0.74	&	0.36	&	0.74 \\
     &  64	&	0.22	&	0.74	&	0.31	&	0.74 \\
$q=8$ &  96	&	0.11	&	0.74	&	0.12	&	0.74 \\
     &  128	&	0.074	&	0.74	&	0.09	&	0.73 \\
     &  192	&	0.045	&	0.73	&	0.052	&	0.73 \\
     &  256	&	0.032	&	0.73	&	0.037	&	0.73 \\
\hline
     &  16	&	0.74	&	0.71	&	0.50	&	0.71 \\
     &  32	&	0.38	&	0.70	&	0.43	&	0.70 \\
$q=10$ &  64	&	0.20	&	0.69	&	0.27	&	0.69 \\
     &  128	&	0.074	&	0.69	&	0.088	&	0.68 \\
     &  256	&	0.032	&	0.69	&	0.026	&	0.68 \\
\hline
     &  16	&	0.84	&	0.57	&	0.64	&	0.57 \\
     &  32	&	0.57	&	0.57	&	0.48	&	0.57 \\
$q=20$ &  64	&	0.18	&	0.57	&	0.17	&	0.56 \\
     &  128	&	0.07	&	0.57	&	0.06	&	0.52 \\
     &  256	&	0.029	&	0.57	&	0.018	&	0.52 \\
\hline
\end{tabular}
\label{tabI}
\end{table}

\section{Hysteresis}
\label{sec:hysteresis}

One of the most characteristic features of first-order transitions is the occurrence
of metastability, i.e., the system remains in one phase when the transition point is
crossed even though the free energy of the other phase is lower there. The metastable
states decay to the stable phases subject to perturbations on a time scale that
depends on the cooling (or heating) rate. Only if one moves too far into the opposite
phase regime, metastability disappears \cite{binder:87}. To clearly reveal this
effect in the present setup, we need to cross the phase boundary in both
directions. This is possible through complementing the {\em cooling\/} run used in PA
by an additional {\em heating\/} sweep. The algorithm described above in
Sec.~\ref{sec:algorithm} is in fact independent of the sign of $\Delta\beta$, so a
negative $\Delta\beta$ corresponding to a heating run is a perfectly valid choice.

To fulfill the preconditions of the approach, we only need to make sure that the
starting population, which in contrast to the cooling run is now at the {\em
  lowest\/} temperature, is a well equilibrated sample. If we start runs deep in the
ordered phase, however, this can easily be achieved by simulating the ensemble for a
few sweeps of local updates at this lowest temperature.  Alternatively, one might
directly prepare the population in the ground-state manifold. For the present case,
this corresponds to a uniform distribution of replicas over the $q$ ground states,
i.e.,
\begin{equation}
  \label{eq:uniform}
  s_k = \alpha,\;\;\;k=1,\ldots,N,
\end{equation}
where $\alpha$ is an integer random variable with uniform distribution,
${\mathbb P}(\alpha = j) = 1/q$. This corresponds to an equilibrium sample for
$T=0$. In practice, it is an excellent approximation also for small $T > 0$, and we
start our runs at $\beta = 3.0$.

\begin{figure}[t]
  \begin{center}
    \includegraphics[scale=0.8]{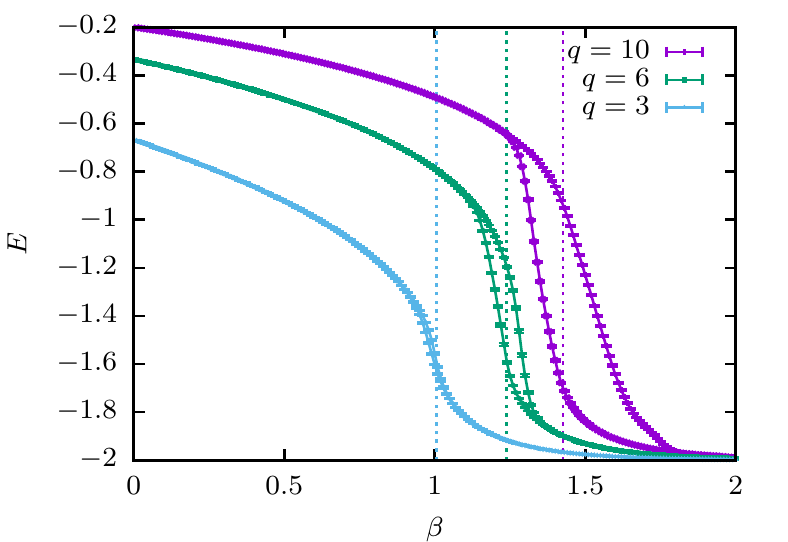}
    \includegraphics[scale=0.8]{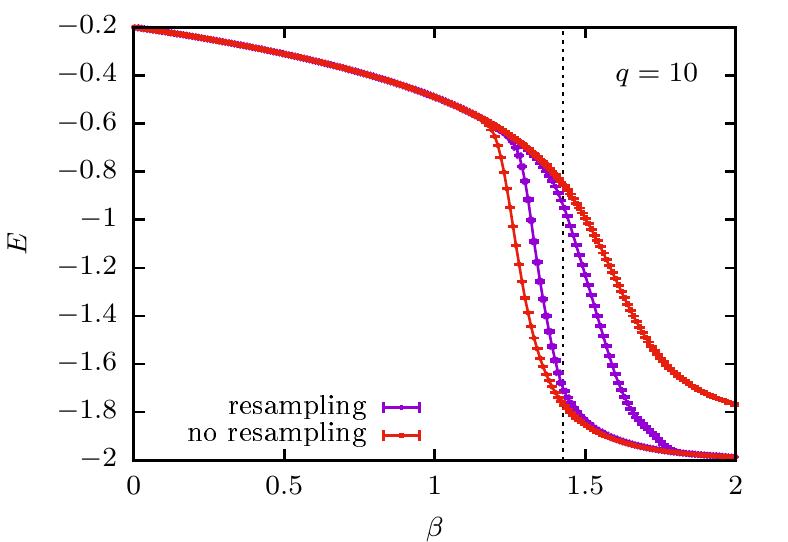}
  \end{center}
  \caption{\label{fig:hysteresis}%
    Left: Internal energy as measured in cooling runs (right curve of each color) and
    heating runs (left curve of each color) for $q=3$, $q=6$ and $q=10$ from PA runs
    for $L=32$ and $R=10\,000$ with $\theta=10$ and $\Delta\beta = 0.01$. It is
    clearly visible that hysteresis occurs in the first-order cases $q=6$ and $q=10$,
    but not for the second-order model with $q=3$. The vertical dashed lines show the
    asymptotic transition points at $\beta_t = \ln(1+\sqrt{q})$. Right: The
    data for $q=10$ as shown on the left compared to the data for equivalent PA runs with the
    resampling step turned off.
  }
\end{figure}

The difference in energies between cooling and heating runs is shown for $q=3$, $q=6$
and $q=10$ and $L=32$ in the left panel of Fig.~\ref{fig:hysteresis}. The heating
runs use the same temperature sequence as the cooling runs (but in reverse
order). While in the second-order regime for $q=3$ the cooling and heating curves
coincide within statistical errors, as expected, this is not the case for the
first-order models with $q=6$ and $q=10$. This hysteresis effect increases with the
strength of the transition and hence with the value of $q$.  As the vertical dashed
lines indicate, the area in the hysteresis loop is approximately, but clearly not
perfectly divided in half by the asymptotic transition line \cite{berg:04a}. We thus
see clearly that PA in its standard setup is not able to equilibrate the population
in the vicinity of the transition point. Still, the resampling strongly reduces the
hysteresis effect, at least for the small system size considered here. This is
illustrated in the right panel of Fig.~\ref{fig:hysteresis}, where we compare PA runs
with and without resampling.

\begin{table}[t]
\caption{%
  Maximal histogram widths and temperatures of maxima for the energy and
  magnetization for the $8$-state Potts model on an $L=64$ lattice with $R=1000$
  replicas as a function of $\theta$. The other parameters were chosen as explained
  in the caption of Table \ref{tabI}.
}
\center
\begin{tabular}{|r|d{1.2}|d{1.2}|d{1.2}|d{1.2}|}
\hline
\multicolumn{1}{|c|}{$\theta$} & \multicolumn{1}{c|}{$\omega_{L,\mathrm{max}}(e)$} &
           \multicolumn{1}{c|}{$T_{\mathrm{max}}(e)$} &
           \multicolumn{1}{c|}{$\omega_{L,\mathrm{max}}(m)$} &
           \multicolumn{1}{c|}{$T_{\mathrm{max}}(m)$} \\
\hline
10	&	0.10	&	0.73	&	0.14	&	0.72 \\
25	&	0.13	&	0.74	&	0.20	&	0.73 \\
50	&	0.17	&	0.74	&	0.23	&	0.74 \\
100	&	0.22	&	0.74	&	0.31	&	0.74 \\
200	&	0.28	&	0.74	&	0.39	&	0.74 \\
500	&	0.29	&	0.74	&	0.36	&	0.74 \\
1000	&	0.34	&	0.74	&	0.42	&	0.74 \\
\hline
\end{tabular}
\label{tabII}
\end{table}

\begin{table}[b]
\caption{%
  Maximal histogram widths and temperatures of maxima for the energy and
  magnetization for the $8$-state Potts model on an $L=64$ lattice with $R=1000$
  replicas and $\theta=100$ sweeps as a function of the number $N_\beta$ of
  temperature steps in the vicinity of the transition coupling $\beta_t$. 
  The temperature protocol is described in detail in the caption of
  Table \ref{tabI}.
}
\center
\begin{tabular}{|r|d{1.2}|d{1.2}|d{1.2}|d{1.2}|}
\hline
\multicolumn{1}{|c|}{$N_\beta$} &
\multicolumn{1}{c|}{$\omega_{L,\mathrm{max}}(e)$} &
\multicolumn{1}{c|}{$T_{\mathrm{max}}(e)$} &
\multicolumn{1}{c|}{$\omega_{L,\mathrm{max}}(m)$} &
\multicolumn{1}{c|}{$T_{\mathrm{max}}(m)$} \\
\hline
100	&	0.11	&	0.73	&	0.15	&	0.72 \\
200	&	0.13	&	0.74	&	0.21	&	0.72 \\
500	&	0.17	&	0.74	&	0.27	&	0.73 \\
1000	&	0.23	&	0.74	&	0.34	&	0.74 \\
2000	&	0.27	&	0.74	&	0.38	&	0.74 \\
\hline
\end{tabular}
\label{tabIII}
\end{table}

To additionally illustrate the hysteresis effect in PA, we produced animations of the
temperature sweeps showing the evolution of a randomly picked replica in the PA
population for the two cases of increasing and decreasing temperatures. These videos
are available in the supplementary material \cite{suppl}. Two videos show annealing
(cooling) of the square lattice of $64\times 64$ spins for $q=6$ (correlation length
$\xi\approx 160$) and for $q=20$ ($\xi\approx 3$). A further two videos show the
heating runs for the same models. As is clearly visible, the ordering and disordering
occurs in a way that is not symmetric between cooling and heating, indicative of the
hysteresis and metastability. Note that the $q=6$ model with its large correlation
length at the transition point shows similarities to the ordering behavior of a
system with a continuous transition as for the given example
$L=64 \ll \xi\approx 160$.

\section{The free energy}
\label{sec:integration}

A classical method for the determination of the phase boundary in first-order
transitions is the comparison of the free energies of the two phases as a function of
the control parameter, here the temperature. The transition occurs where the two
pure-phase branches of the free energy cross \cite{janke:03,binder:book2}. In
standard Monte Carlo simulations it is not straightforward to produce reliable
estimates of the free energy, as it cannot be directly derived from a configurational
observable. The standard approach is through {\em thermodynamic integration\/}, which
relies on the relation $E = \partial (\beta F)/\partial \beta$, such that a numerical
integral of the internal energy over a temperature range will yield an estimate for
the difference of free energies at the endpoints of the interval
\cite{binder:book2}. The absolute normalization is additionally derived from exact
calculations for $Z_{\beta_0}$ as indicated below Eq.~\eqref{eq:free-energy} or
from high- or low-temperature series expansions \cite{janke:03}.

In PA, a reliable estimator of free energies is explicitly provided through the
resampling factors that are combined in the estimator
Eq.~\eqref{eq:free-energy}. For the case of the cooling schedule, we start at
$\beta_0 = 0$ and hence we have $Z_{\beta_0} = q^N$. For the heating runs, on the
other hand, we note that
\begin{equation}
  \label{eq:groundstate}
  Z_{\beta\to\infty} = \lim_{\beta\to\infty} \sum_{\{s_k\}}e^{-\beta{\cal
      H}(\{s_k\})} = \lim_{\beta\to\infty} q\,e^{-\beta E_0},
\end{equation}
where $E_0$ is the ground-state energy that equals  $E_0 = -2N$ for the
square-lattice model with periodic boundaries studied here. The free energy in this
limit hence becomes
\begin{equation}
  \label{eq:free-energy-zeroT}
  -\frac{\beta F_{\beta\to \infty}}{N} = \frac{\ln q}{N} - \beta e_0,
\end{equation}
where $e_0 = E_0/N = -2$.

\begin{figure}[t]
  \begin{center}
    \includegraphics[scale=0.8]{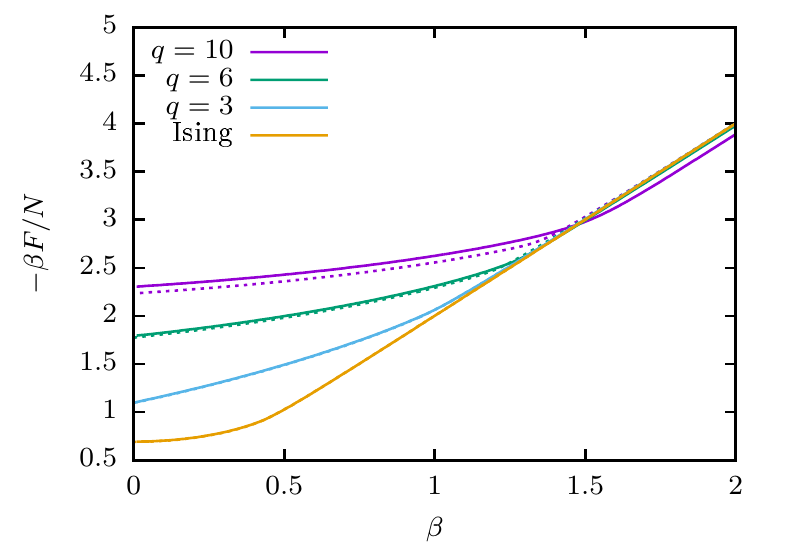}
    \includegraphics[scale=0.8]{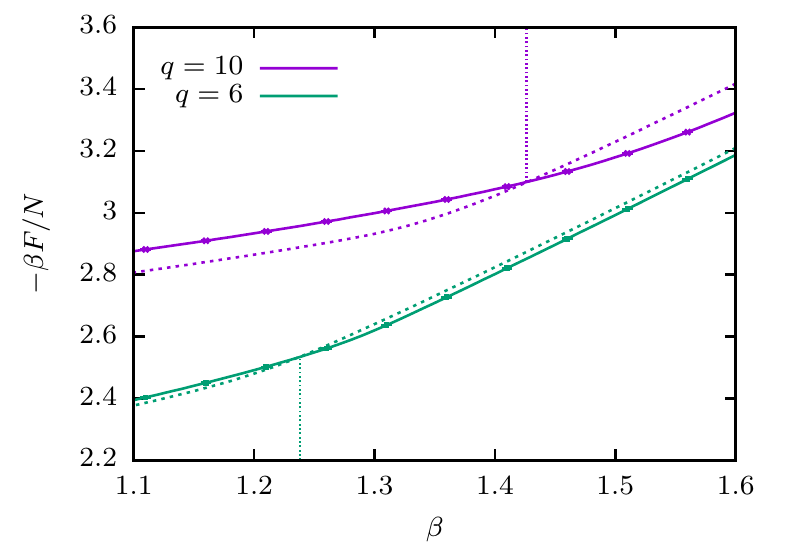}
  \end{center}
  \caption{\label{fig:free-energy}%
    Left: Free energy of the Ising model and the $q=3$, $q=6$ and $q=10$ Potts models
    as estimated from PA runs with cooling (solid lines) and with heating (dashed
    lines). Parameters are $L=32$, $R=10\,000$, $\Delta\beta = 0.01$ and
    $\theta = 10$. For the Ising model and $q=3$, the two estimates coincide
    everywhere, whereas for the first-order cases $q=6$ and $q=10$ each estimate
    ceases to correspond to the equilibrium free energy as soon as the transition
    point is crossed. Right: Detail of the crossing of the metastable free energies
    for $q=6$ and $q=10$. Symbols with error bars are only shown for
    every fifth actual data point. The data for $q=10$ have been shifted vertically
    for clarity of presentation. The vertical dotted lines indicate the locations
    of the asymptotic transition points $\beta_t = \ln(1+\sqrt{q})$.
  }
\end{figure}

In Fig.~\ref{fig:free-energy} we show the resulting free-energy estimates from
cooling runs starting from $\beta_0 = 0$ as compared to heating runs started from an
equidistribution in the ground states of the system at the initial inverse
temperature $\beta_{K-1} = 3$ and using the normalizations resulting from
$Z_{\beta_0}$ and $Z_{\beta\to\infty}$. As is seen in the left panel, the two
estimates coincide everywhere for the second-order cases of the Ising model
(corresponding to $q=2$, but with a different normalization) and the $q=3$ Potts
model, but differences appear for the first-order systems $q=6$ and $q=10$. As the
right panel reveals, the two metastable free energies cross very close to the
asymptotic critical point. From the simulations with $\Delta\beta = 0.01$ we can only
determine the crossing points with a resolution of $\Delta\beta$, and the locations
of the crossings are consistent with the asymptotic $\beta_t = \ln(1+\sqrt{q})$
already for $L=32$ studied here. One can easily imagine using a finer temperature
grid in the relevant temperature regime to improve on these results.

\section{Conclusion}
\label{sec:conclusions}

We have presented a preliminary report on the behavior of the population annealing
algorithm when applied to a system with a first-order phase transition. As a well
understood example system we considered the Potts model on the square lattice with
$q > 4$ states. While the resampling element reduces the effect of metastability and
hysteresis, it is not able to remove it, at least without further modifications of
the algorithm. Still, the possibility of reliably estimating free energies turns out
to be a useful feature of the method also for the study of systems with first-order
transitions as it appears to allow for a reasonably precise estimate of the
transition point through the matching of the pure-phase free-energy branches. While
population annealing in the present setup does not appear to be an ideal tool
for systems with discontinuous transitions, the search for a variant of the approach
using a modified ensemble and possibly modified update moves promises some
improvement in this respect.

\acknowledgement{ The article is dedicated to Wolfhard Janke on the occasion of his
  60th birthday. The authors acknowledge support from the European Commission through
  the IRSES network DIONICOS under Contract No. PIRSES-GA-2013-612707. L.B.\ and
  L.S.\ were supported by the grant 14-21-00158 from the Russian Science Foundation.
  M.W.\ is grateful to Coventry University for providing a Research Sabbatical
  Fellowship that supported a long-term visit to Leipzig University, where part of
  this work was performed.
}

\end{document}